\documentclass[slac_one]{revtex4}
\usepackage{graphicx}
\usepackage{fancyhdr}
\usepackage{amsfonts}
\usepackage{mathrsfs}
\newcommand{\fb}{\ensuremath{{\rm fb}^{-1}}}

\newcommand{\lumi}{\ensuremath{\mathscr{L}}}

\newcommand{\instlumi}[1]{\ensuremath{\lumi=10^{#1}$\,cm$^{-2}$s$^{-1}}}

\newcommand{\Mhalf}{\ensuremath{M_{1\!/\!2}}}

\newcommand{\mll}{\ensuremath{M_{\ell\ell}}}
\newcommand{\mossf}{\ensuremath{\mll^{\rm OSSF}}}
\newcommand{\etal}{\textit{et al.}}
\newcommand{\slashchar}[1]{\ensuremath{/\!\!\!\!{#1}}}           
\newcommand{\met}{\ensuremath{E_T^{\rm miss}}}
\newcommand{\metslash}{\ensuremath{\slashchar{E}_{T}}}
\newcommand{\mpt}{\ensuremath{\slashchar{p}_{T}}}
\newcommand{\ttbar}{\ensuremath{t\bar t}}
\newcommand{\meff}{\ensuremath{M_{\rm eff}}}

\begin{document}

\title{{\small{Hadron Collider Physics Symposium (HCP2008),
Galena, Illinois, USA}}\\ 
\vspace{12pt}
Supersymmetry Searches at the LHC} 

\author{O.~Brandt on behalf of the ATLAS and CMS collaborations}
\affiliation{University of Oxford, Keble Road, Oxford, OX1 3RH, UK}

\begin{abstract}
The search for Beyond the Standard Model (BSM) physics is one of the major tasks of the LHC, CERN. In these proceedings, I review the status of searches for Supersymmetry by the ATLAS and CMS collaborations. The efforts in both the hadronic and leptonic search channels are presented. A special focus is placed on the treatment of approximately 1\,\fb\ of early LHC data, and examples of background estimation techniques in such a dataset are given. Phenomenologically, besides ``typical'' mSUGRA scenarios, signatures based on prompt and non-pointing photons, as well as long-lived leptons and hadrons ($R$-hadrons) are covered.
\end{abstract}

\maketitle

\thispagestyle{fancy}

\section{Introduction}
One of the main goals of the ATLAS and CMS general purpose detectors at the Large Hadron Collider (LHC)~\cite{bib:pdg}, a proton-proton collider with a design centre-of-mass energy of $\sqrt s = 14$ TeV, is to search for new physics beyond the Standard Model (BSM). A widely favoured BSM candidate is Supersymmetry (SUSY), which will be introduced in Subsection~\ref{ssec:susy}.

Several strategies can be applied to search for \textit{direct} experimental evidence of SUSY at a collider experiment. 
In these proceedings, I will present a wide cross-section through SUSY search analyses at ATLAS and CMS, grouped by lepton multiplicity, since each of these channels involves a specific class of backgrounds. A special focus will be placed on techniques to measure background contributions from data.

\subsection{The ATLAS and CMS detectors}
ATLAS and CMS detectors are the two major general purpose experiments at the LHC. Both can be characterised as classical collider detectors with full solid angle coverage. 

Major components of ATLAS~\cite{bib:tdr} are the Inner Detector, consisting of the Pixel Detector, the SemiConductor Tracker and the Transition Radiation Tracker; the liquid argon electromagnetic and the hadronic calorimeters; as well as the Muon Spectrometer. The Inner Detector is situated inside of a 2\,T solenoidal magnetic field, whereas the Muon Spectrometer features a toroidal magnetic field with a bending power of 1.5 to 5\,Tm in the barrel~\cite{bib:detectorPaper}.

The most striking feature of the CMS detector~\cite{bib:tdrCMS} is that the Inner Tracker and the calorimetry are immersed inside of a compact solenoidal 4\,T magnetic field, which reverses its direction in the Muon Spectrometer. The main components of CMS are: the Pixel and Semi-Conducting Strip detector, the lead tungstate electromagnetic calorimeter, the hadronic calorimeter, and the Muon Spectrometer.

\subsection{Supersymmetry}\label{ssec:susy}
The Standard Model of elementary particle physics is a relativistically and local gauge invariant, renormalisable quantum field theory of strong and electroweak interactions. In recent decades, it has proven extremely successful in explaining the observed phenomena. However, evidence is mounting~\cite{bib:wmap,bib:ellisOlive} that it is merely a low-energy approximation of another theory.
One of the major problems of the Standard Model is arguably the hierarchy problem~\cite{bib:susy}. An elegant and \ae{}sthetic solution 
is offered by Supersymmetry (SUSY)~\cite{bib:susy,bib:susyHCP}, which for each particle of the Standard Model postulates a supersymmetric partner with the same quantum numbers, but with spin quantum number different by~1/2. This naturally solves the hierarchy problem. In this report, I will only focus on the the simplest supersymmetric extension of the Standard Model, the Minimal Supersymmetric Standard Model (MSSM)~\cite{bib:susy}.

An elegant mechanism for spontaneous soft supersymmetry breaking required for theoretical sanity is proposed by Supergravity (SUGRA). It postulates SUSY breaking  in a ``hidden'' sector at a scale of around 10$^{10}$\,GeV, which is mediated to the ``visible'' sector by flavour-blind gravitational interactions. Another well-known mechanism is Gauge Mediated Symmetry Breaking (GMSB), where the breaking happens at a much lower scale and is mediated via an intermediate messenger sector by $\mathbb{SU}(3)_c\times\mathbb{SU}(2)_Y\times\mathbb{U}(1)_{em}$ interactions.

\section{Generic SUSY Search Signatures at the LHC}\label{sec:generic}
Lepton and baryon quantum number conservation provides a strong phenomenoligical constraint on any extension to the Standard Model, since we know experimentally that the lifetime of the proton is longer than $10^{31}$\,s~\cite{bib:pdg}. This property is preserved if absolute $R$-parity conservation is imposed~\cite{bib:susy}. Consequences of R-parity conservation are that particles must be pair-produced and that the Lightest Super\-symmetric Particle (LSP) is stable, which makes it, if not electromagnetically or strongly charged, a perfect candidate for dark matter~\cite{bib:darkmatter}. The resulting generic experimental signature is characterised by two decay chains of supersymmetric particles, and significant $\met$ values from the LSPs escaping detection. Both ATLAS and CMS have developed search analyses based on these generic signatures, while significant efforts have been made to avoid specialisation towards any particular point in the mSUGRA phase space. I will describe some of these searches in this Section, ordered by lepton multiplicity. 
All the figures presented below are for 1\,\fb\ unless stated otherwise.

\subsection{The mSUGRA Parameter Space and Benchmark Points at ATLAS and CMS}\label{sssec:0l_trijet}
\begin{figure}
\begin{center}
\begin{picture}(175,81)
\put(10,0){
\includegraphics[width=5.18cm]{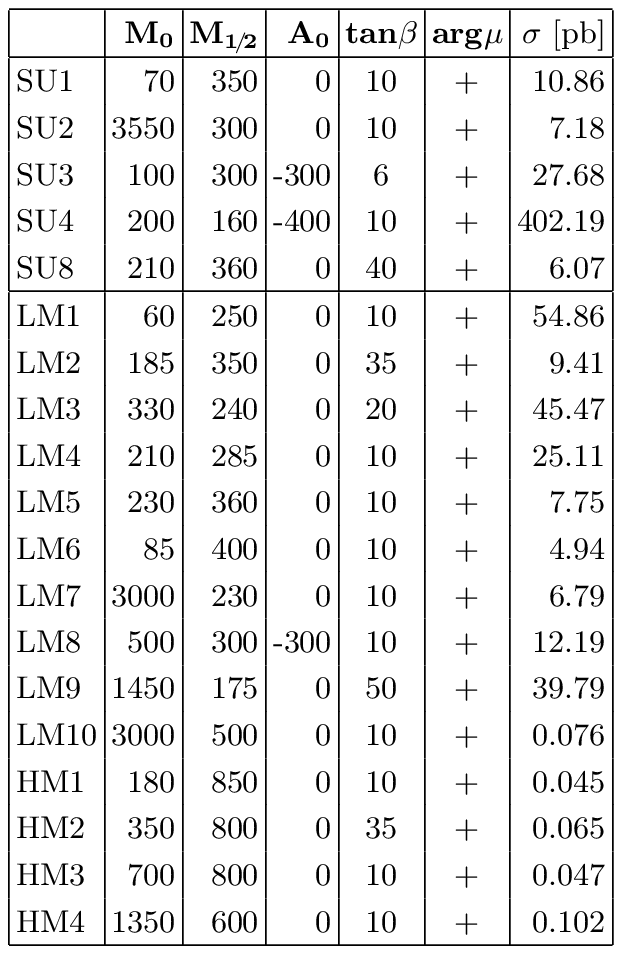}
}
\put(75,0){
\includegraphics[width=9cm,height=8.12cm]{figures/ptdr1_final_colour.eps2}
}
\put(5,80) {($a$)}
\put(75,80) {($b$)}
\end{picture}
\caption{\label{fig:mSUGRAPoints}
mSUGRA parameter values for the benchmark points used by the ATLAS (SU$x$, in red) and CMS (LM$x$, HM$x$, in blue) collaborations together with the resulting cross-sections at NLO calculated with Prospino~\cite{bib:prospino} are shown in ($a$). The $M_0,\,\Mhalf,\,A_0$ parameters are given in GeV. The positions of the benchmark points in the plane of $M_0$ and \Mhalf, which govern the mass scales of supersymmetric scalars and gauginos, respectively, can be seen in~($b$).
}
\end{center}
\end{figure}

The SUGRA parameter space is much too large to be experimentally investigated in its entirety. Experimentally it is convenient to explore a phenomenologically simplified model, the minimal version of Supergravity (mSUGRA), which is governed by only 5 parameters at GUT scale~\cite{bib:susy}. Although mSUGRA imposes particular mass relationships, its phase space is expected to cover a wide variety of signature types expected in SUGRA and to some extent in MSSM. For detailed simulation studies, both ATLAS and CMS collaborations have defined a set of representative benchmark points in the mSUGRA parameter space. They are summarised in Figure~\ref{fig:mSUGRAPoints}.

\subsection{No-lepton Mode}
Since the LHC is a proton-proton machine, one can expect that the production of supersymmetric particles will be dominated by strong processes. Assuming this and moderate relative branching ratios to leptonic modes, the no-lepton search mode will account for the dominating fraction of the total SUSY cross-section. However, this channel is characterised by high backgrounds from QCD. A requirement of high \met{} values is very effective in rejecting this background. Given that production of strongly interacting particles dominates the cross-section, a thorough understanding of instrumental effects which are accountable for the non-Gaussian tails in the \met{} distribution is necessary. Othemportant backgrounds are the $\ttbar$ production, $Z+j\rightarrow\nu\bar\nu+j$ processes, as well as $W+j$, where the lepton from $W$ decay is missed. Respecting the importance of this channel, two analyses, one from ATLAS and one from CMS, shall be presented.

\subsubsection{No-lepton Mode Trijet Search at CMS}
The no-lepton mode analysis of CMS~\cite{bib:physicsTDR} is built around the requirement of high missing transverse energy and at least three jets in the event. Accordingly, at level-1 a jet trigger with $E_T^j>88$\,GeV AND-combined with $\met>46$\,GeV is required. The high level trigger demands only the $\met>200$\,GeV criterion. This configuration is expected to give robust results during the early data taking period.

\begin{figure}
\begin{center}
\begin{picture}(175,54)
\put(0,0){
\includegraphics[width=8.5cm,height=5.4cm,clip]{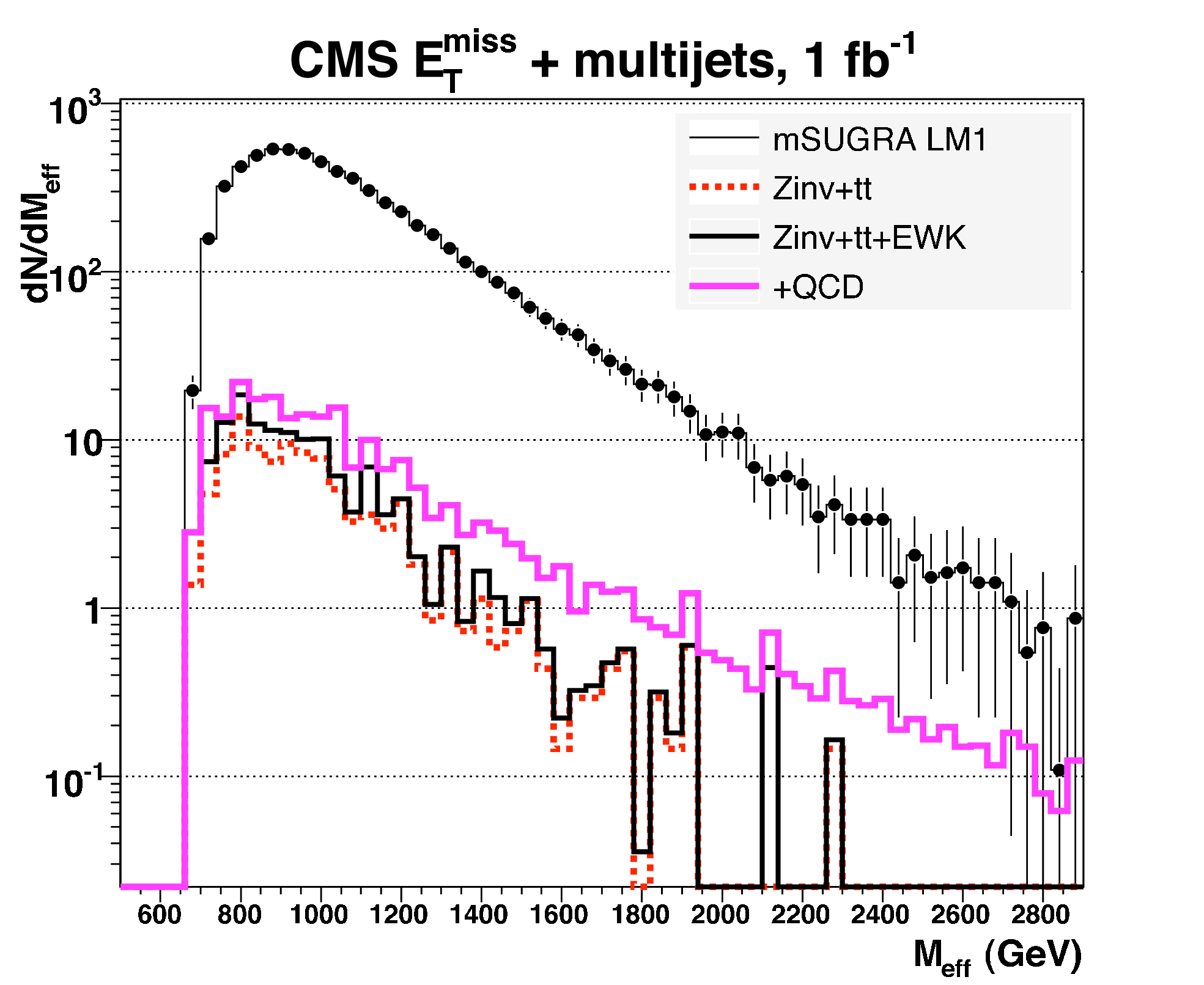}
}
\put(85,1){
\includegraphics[width=8.5cm,height=5.3cm,clip]{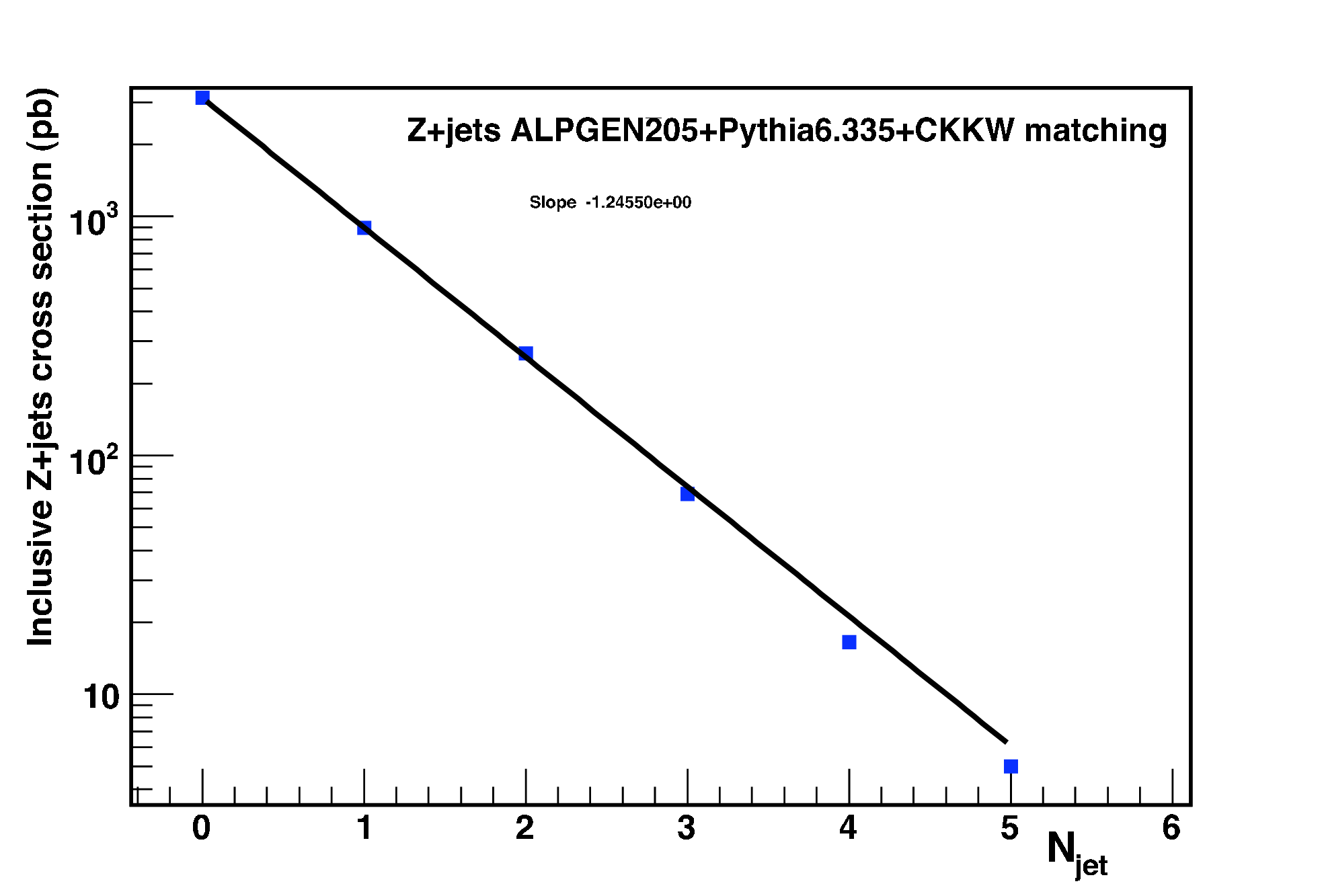}
}
\put(12,48) {($a$)}
\put(97,48) {($b$)}
\put(70,35) {{\sf\large CMS}}
\put(155,35) {{\sf\large CMS}}
\end{picture}
\caption{\label{fig:0leptCMS}
($a$)~Effective mass \meff\ distribution for the CMS no-lepton trijet search after all cuts. ($b$)~Illustration of $dN_{\rm event}/dN_j$ measurement in $Z+j$ pseudo data used to estimate the $Z\rightarrow\nu\bar\nu$ background for the same analysis.
}
\end{center}
\end{figure}
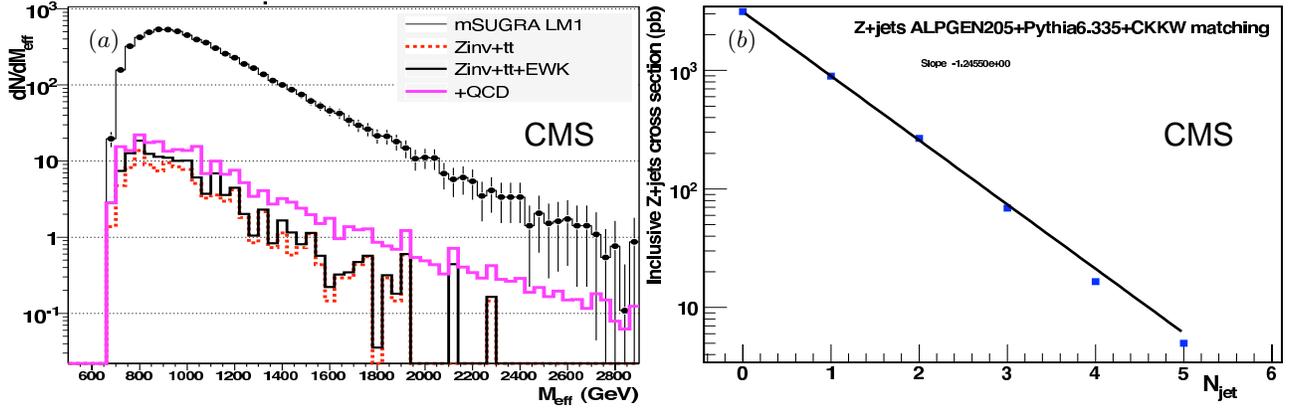

\pagebreak[3]
In the offline selection, three high-$E_T$ jets are required, of which one needs to be central:
\begin{itemize}
\vspace{-2mm}
  \item {\bf 1$^{\rm st}$ jet:} $E_T^{j_1}>180$\,GeV, $|\eta^{j_1}|<1.7$;
  \vspace{-2mm}
  \item {\bf 2$^{\rm nd}$ jet:} $E_T^{j_2}>110$\,GeV, $|\eta^{j_2}|<3$;
  \vspace{-2mm}
  \item {\bf 3$^{\rm rd}$ jet:} $E_T^{j_3}>30$\,GeV, $|\eta^{j_3}|<3$.
\vspace{-2mm}
\end{itemize}
Besides the three jets requirement a cut $\met>200$\,GeV is applied. 
In QCD background events, such high \met\ values can come about if one or more of the jets are mismeasured. In such a case, the \met\ vector will tend to be aligned with the vector of the most mismeasured jet with the highest $E_T$. It is therefore required that the azimuthal angle between the \met\ vector and any jet with $E_T>30$\,GeV is greater than 0.3. It is also required that there be a reasonable amount of hadronic activity in the event: $H_T\equiv{\sum}_{k=2,3,4}E_T^{j_k}+\met>500$\,GeV.

In order to suppress backgrounds from electroweak processes and semileptonic \ttbar, an implicit lepton veto is made: all events with an isolated track with $p_T^{\rm track}>15$\,GeV are discarded, as are events where the electromagnetic fraction of the two leading jets is higher than 0.9. This results in the effective mass distribution shown in Figure~\ref{fig:0leptCMS}~($a$).

There are several studies at CMS aiming at estimating background contribution from data. As an example, I will present an approach to measure the contribution of the $Z+3j\rightarrow\nu\bar\nu+3j$ background. The basic idea is to take a similar process, $Z+3j\rightarrow\mu\bar\mu+3j$ as a control sample, and then replace the muons with neutrinos. This implies setting $p_T^Z(\mu\bar\mu)\equiv\met(\nu\bar\nu)$. 
However, this approach inherently suffers from low statistics of the control sample because $BR(Z\rightarrow\mu\bar\mu)/BR(Z\rightarrow\nu\bar\nu)\simeq0.17$. To increase the number of events in the control region, a ``bootstrap'' method is used: the ratio of $Z+2j$ to $Z+3j$ processes scales like $\alpha_s$ and can be measured, as indicated in Figure~\ref{fig:0leptCMS}~($b$). Therefore, this background contribution can be approximated by taking Monte Carlo-simulated events with $N_j\geq3$ and normalising them to the $Z+2j\rightarrow\mu\bar\mu+2j$ data sample. In a similar fashion, the $W+2j\rightarrow\tau\nu+2j$ with hadronic $\tau$~decays can be estimated, using the ratio of the $W$ and $Z$ production cross-sections. The drawback of this method is that despite using the $N_j\geq2$ jet bin it requires approximately 1\,\fb\ of integrated luminosity.

\subsubsection{No-lepton Mode Dijet Search at ATLAS}\label{sssec:0l_dijet}
In the ATLAS TDR a no-lepton analysis which required four jets was studied~\cite{bib:tdr}. Selections requiring smaller jet multiplicities are also potential early SUSY-search channels, and here I present a new analysis which only requires two jets~\cite{bib:csc}. It uses a new discrimination variable $m_{T2}$~\cite{bib:mt2}, which is similar to the $m_T$ variable for $W$ mass measurement. The definition of $m_{T2}$ is:
\begin{equation}
 m_{T2}^2({\bf p}_T^\alpha, {\bf p}_T^\beta,  {\bf \mpt} ,
 m_\alpha, m_\beta, m_\chi)
 \equiv { \min_{\slashchar{{\bf q}}_T^{(1)} +
 \slashchar{{\bf q}}_T^{(2)} = {\bf \mpt} }} {\Bigl[ \max{ \Bigl\{
 m_T^2({\bf p}_T^\alpha, \slashchar{{\bf q}}_T^{(1)}; m_\alpha, m_\chi) ,\
 m_T^2({\bf p}_T^\beta,  \slashchar{{\bf q}}_T^{(2)}; m_\beta, m_\chi)
\Bigr\} }
\Bigr]}\,.\label{eqn:mt2}
\end{equation}
This variable has interesting properties, namely it takes low values for events where either the visible jet $p_T$ or the $\met$ are small, or in case of small azimuthal angle between them, and thus effectively replaces topological cuts. For semi-invisibly decaying particles $m_{T2}$ is correlated with the difference in mass between the particles produced in the interactions and and their invisible decay products. Thus, it is expected to take higher values for SUSY events than for \ttbar{} or $W$ events. The strategy of this search is to require a certain value for $m_{T2}$ and few other kinematic cuts.

\begin{figure}
\begin{center}
\begin{picture}(175,54)
\put(0,0){
\includegraphics[width=8.5cm,height=5.4cm]{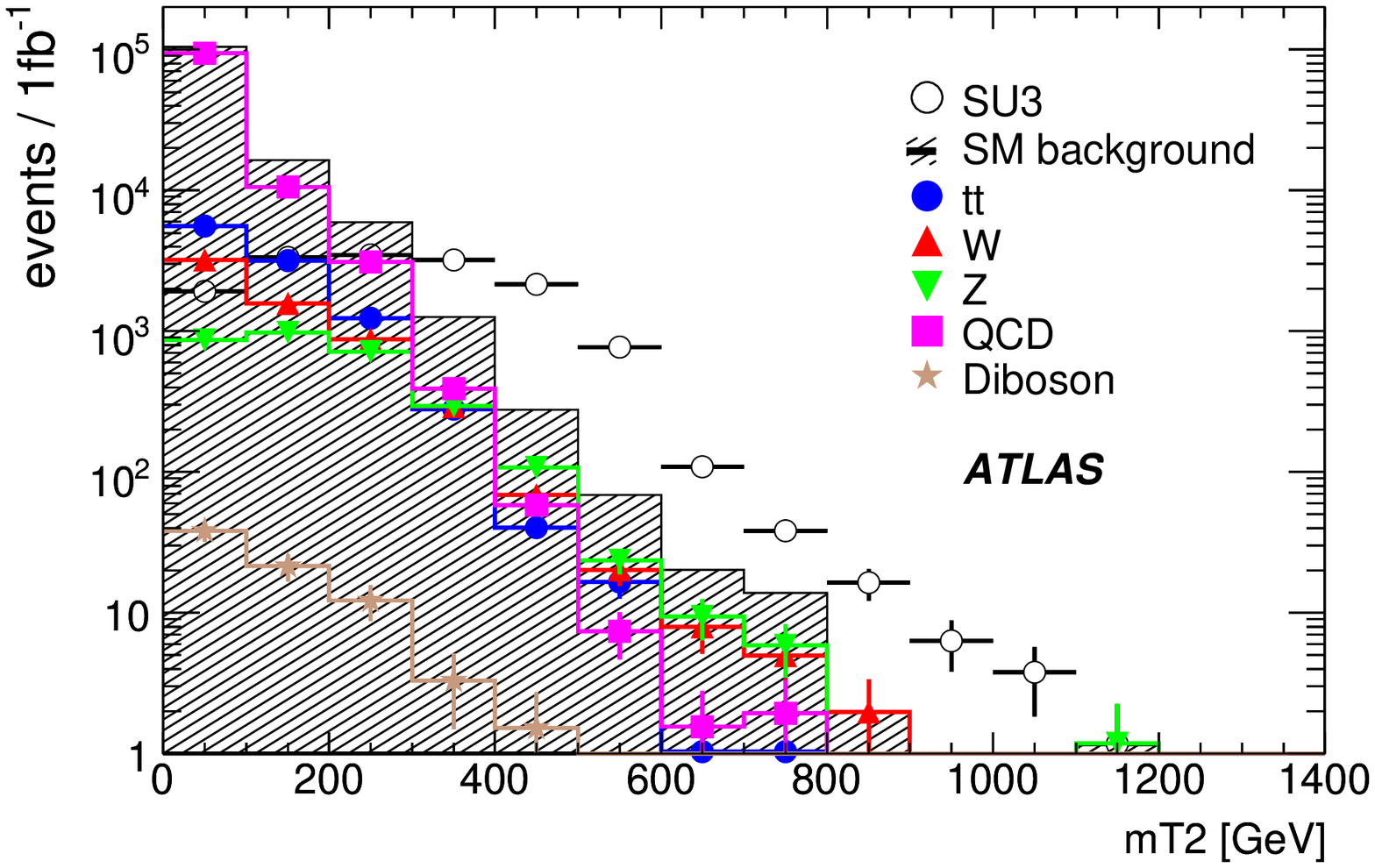}
}
\put(85,0){
\includegraphics[width=8.5cm,height=5.4cm]{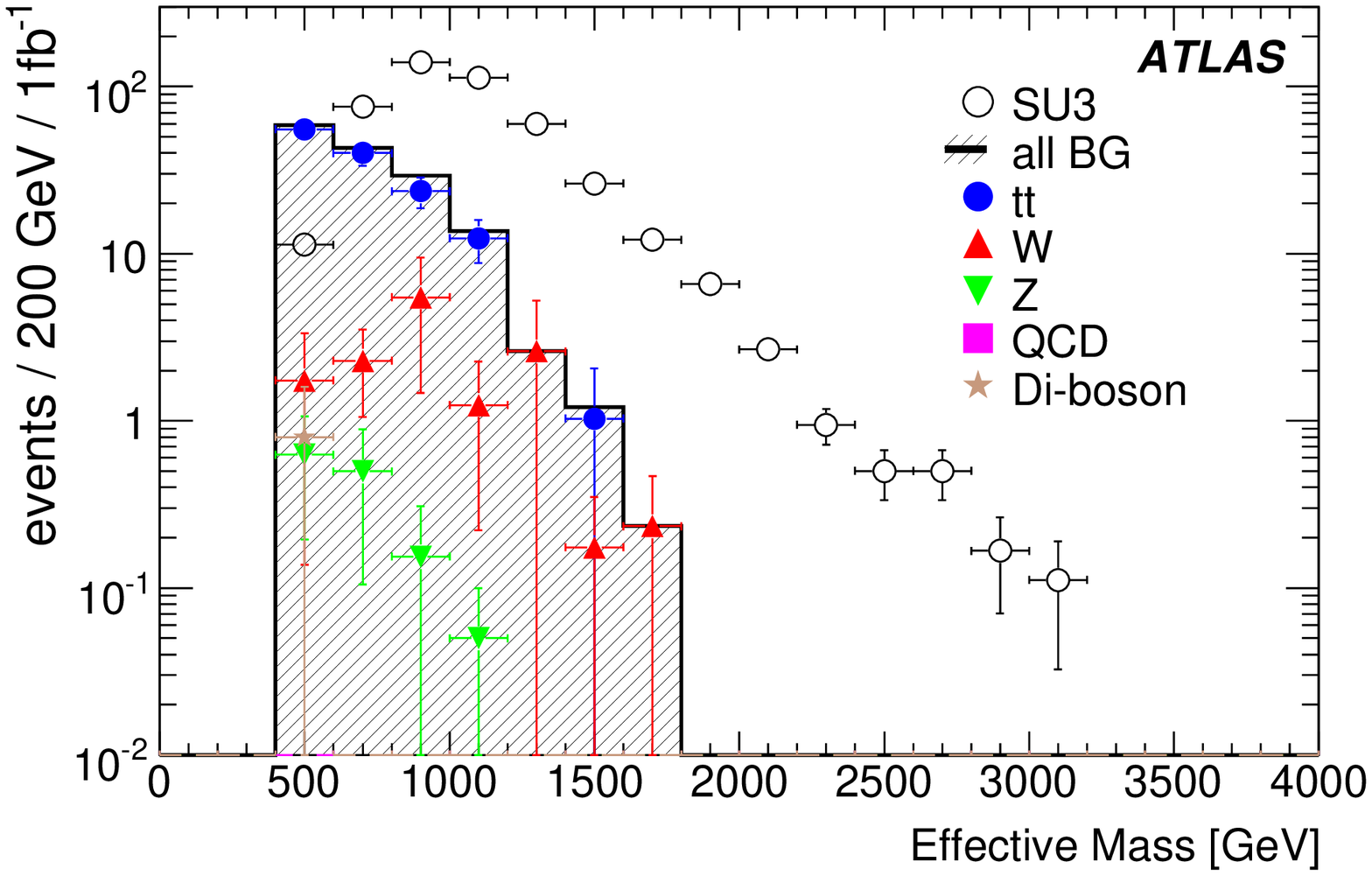}
}
\put(22,48) {($a$)}
\put(100,48) {($b$)}
\end{picture}
\caption{\label{fig:0leptATLAS}
($a$)~Distribution of $m_{T2}$ for the ATLAS no-lepton dijet search after all cuts except those on $m_{T2}$ itself. ($b$)~Effective mass distribution for the ATLAS one-lepton + four jets analysis after all cuts except those on \meff\ itself.
}
\end{center}
\end{figure}

The dijet search uses a rather typical SUSY trigger: $E_T^j>70$\,GeV and $\met>70$\,GeV are required. While reasonably inclusive, it is expected to be unprescaled up to $\instlumi{33}$~\cite{bib:csc}. 
The offline requirements of this analysis are: two jets in $|\eta|<2.5$ with $p_T^{j_1}>150$\,GeV and $p_T^{j_2}>100$\,GeV, as well as $\met>100$\,GeV. In order to enforce orthogonality to other lepton multiplicity channels, a veto on any isolated leptons in the event with $p_T>10$\,GeV is imposed. At the final cut stage $m_{T2}>400$\,GeV is required. The corresponding $m_{T2}$ distribution before this cut is shown in Figure~\ref{fig:0leptATLAS}~($a$).

\subsection{One-lepton Mode + Four-Jets Search from ATLAS}
The one-lepton mode is expected to have a reach comparable to the no-lepton mode despite smaller statistics. This is because the main backgrounds, QCD and invisibly decaying $Z$ bosons, are much reduced by the requirement of one high-$p_T$ isolated lepton. The main backgrounds for this important channel are semileptonic \ttbar\ decays and associated $W+j$ production. $Z$, diboson and QCD multi-jet production play minor roles.

This analysis uses the same jet+\met\ trigger as the no-lepton dijet search described in Subsection~\ref{sssec:0l_dijet} which is more than $99$\% efficient, rather than a single lepton trigger. Four jets are required offline in $|\eta|<2.5$, with $p_T^{j_1}>100$\,GeV for the leading jet and $p_T^{j}>50$\,GeV for the other three. Requiring a large jet multiplicity is an efficient way of reducing the Standard Model backgrounds. Multi-jet final states are expected to be common in supersymmetric events due to emission of quarks and gluons during cascade decays.

Exactly one lepton ($e$ or $\mu$) in $|\eta|<2.5$ with $p_T>20$\,GeV is required. Events in which a loosely-identified electron is found in the poorly-instrumented barrel-endcap transition region ($1.37<|\eta|<1.52$) are vetoed. To reduce backgrounds from leptons from heavy flavour decays, leptons are required to be isolated -- they must have less than 10 GeV of calorimeter energy inside a $\Delta R<0.2$ cone (in $\eta\times\phi$) around the lepton. 

Beyond the above preselection cuts, this analysis requires:
\begin{itemize}
\vspace{-2mm}
 \item $\met>100$\,GeV;
 \vspace{-2mm}
 \item a big amount of hadronic and leptonic activity: $\meff\equiv{\sum}_kp_T^{j_k}+{\sum}_kp_T^{\ell_k}+\met>800$\,GeV;
 \vspace{-2mm}
 \item a significant fraction of hadronic and leptonic activity due to escaping LSPs: $\met>0.2\meff$;
 \vspace{-2mm}
 \item $S_T\equiv2\lambda_2/(\lambda_1+\lambda_2)>0.2$ in order to select spherical events over the boosted ones. Here, $\lambda_k$ are the $k$-th eigenvalues of the $2\times2$ sphericity tensor $S_{ij}\equiv{\sum}_\ell\, p_{\ell i}p^{\ell j}$, $\ell$ running over all the jets with $p_T>20$\,GeV in $|\eta|<2.5$, and leptons;
 \vspace{-2mm}
 \item $M_T>100$\,GeV in order to reject the $W+j$ background and semileptonic \ttbar\ decays.
\vspace{-2mm}
\end{itemize}
The above requirements result in $\meff$ distribution as shown in Figure~\ref{fig:0leptATLAS}~($b$).

\begin{figure}
\begin{center}
\begin{picture}(175,54)
\put(0,0){
\includegraphics[width=8.5cm,height=5.4cm]{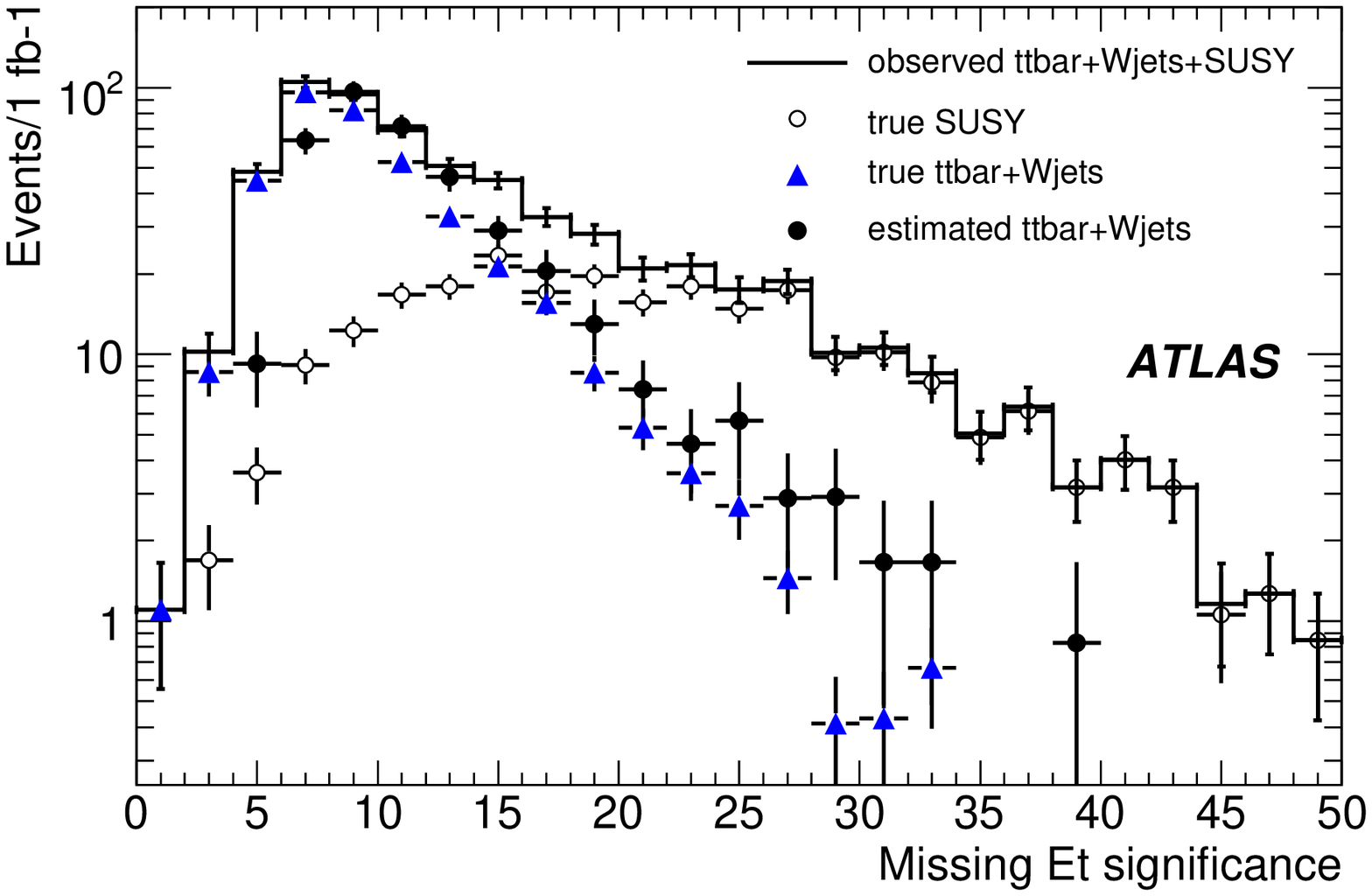}
}
\put(85,0){
\includegraphics[width=8.5cm,height=5.4cm]{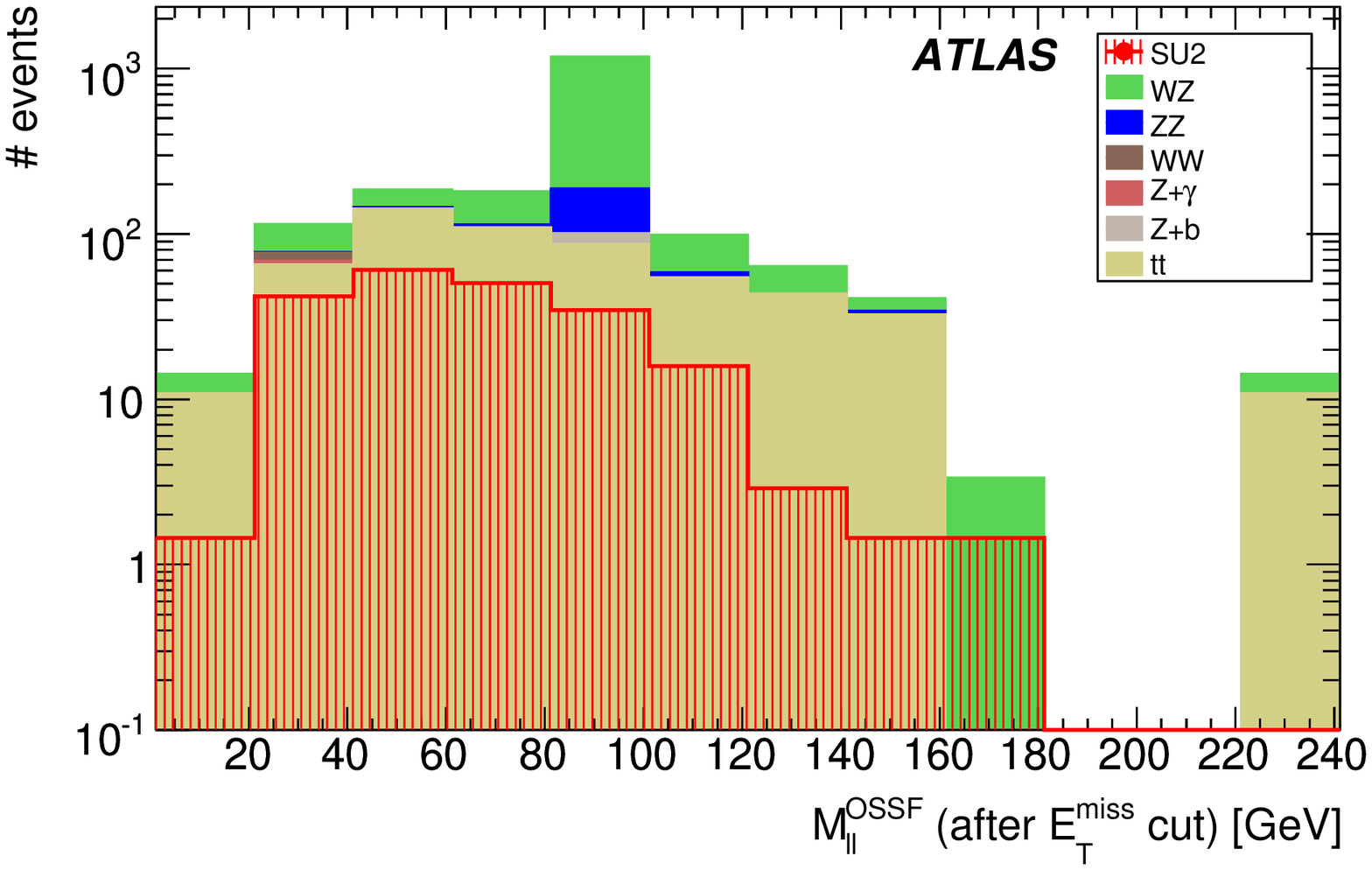}
}
\put(14,46) {($a$)}
\put(99,46) {($b$)}
\end{picture}
\caption{\label{fig:1l_bgr}
($a$)~Comparison of the missing transverse energy significance $\metslash^{\rm sign.}$ distribution for the ATLAS one-lepton search as estimated versus truth. ($b$)~$M_{\ell\ell}$ distribution after all cuts but on $M_{\ell\ell}$ for the ATLAS three-lepton mode search.
}
\end{center}
\end{figure}

ATLAS has developed a wide variety of strategies to estimate backgrounds from data. A rather new method is presented as an example in the following. The $\met$ distribution is among the most difficult ones to model in simulations. At the same time it is a vital part of the above analysis.
This method aims at describing the \met\ distribution for the \ttbar\ and $W+j$ backgrounds from data using a control region. For this, two almost completely uncorrelated variables, $HT2\equiv{\sum}_{k=2,3,4}\,p_T^{j_k}+p_T^\ell$ and $\metslash^{\rm sign.}\equiv\met/(0.49\cdot\sqrt{{\sum}E_T})$, where ${\sum}E_T$ is the total hadronic activity in an event, are chosen. The control sample is defined by $HT2<300$\,GeV, which roughly corresponds to $\meff<600$\,GeV, and is therefore well separated from the signal region. A comparison between the resulting estimated $\metslash^{\rm sign.}$ distribution and the true one in the presence of SU3 SUSY signal is demonstrated in Figure~\ref{fig:1l_bgr}~($a$).

\subsection{Two-lepton Mode Same Sign Muons and Jets Search at CMS}
Same-sign leptons are produced in many SUSY models. This is a striking signature, as irreducible Standard Model backgrounds are small. However, a detailed understanding of the detector is required since instrumental backgrounds will dominate under such conditions.

The CMS analysis~\cite{bib:physicsTDR} presented here investigates only same-sign muon pairs, since their charge mismeasurement rate is expected to be $\mathcal O(100)$ smaller than for electrons in the same kinematical range. A pair of muons satisfying $p_T>7$\,GeV each is demanded in the high level trigger. In the offline selection, their transverse momentum thresholds are raised to 10\,GeV, and same charge is required. Given this very low $p_T$ threshold, high backgrounds from secondary leptons from heavy flavour decays are involved. Therefore, strict isolation criteria both in the tracker and the calorimeter are imposed, which are described in detail in~\cite{bib:physicsTDR}. Besides the leptonic cuts, three jets in $|\eta|<3$ with $E_T^{j_1}>175$\,GeV, $E_T^{j_2}>130$\,GeV, and $E_T^{j_3}>55$\,GeV are required. Finally, a cut on $\met>200$\,GeV is made. For the ``light'' benchmark point LM1, this selection results in 341 signal over 1.5 background events for 10\,\fb, and thus $S/B$ of $\mathcal O(100)$. This clean signal might be used for property measurements of the underlying SUSY model.

\subsection{Three-lepton Mode Search at ATLAS}
Depending on how SUSY is broken, it might happen that masses of squarks and sleptons are much higher than these of gauginos, e.g. due to $M_0\gg\Mhalf$ in mSUGRA, as is the case with the points SU2, LM7, LM9. This will result in gaugino pair production dominating over other SUSY processes. Alternatively, one can think of a scenario where $M_{\rm squarks}\gg M_{\rm sleptons}$. In such cases leptonic modes are likely to present the best discovery potential.

The ATLAS search analysis~\cite{bib:csc} in the three-lepton channel requires the OR-combination of an isolated electron with $p_T^e>25$\,GeV or a muon with $p_T^\mu>20$\,GeV. Preliminary studies show that these triggers can be run unprescaled up to $\instlumi{33}$. The offline threshold is $p_T>10$\,GeV for both $e$ and $\mu$. In order to suppress backgrounds from leptons from heavy flavour decays, the leptons must have less than 10\,GeV of calorimeter energy inside a cone in $\eta\times\phi$ around the lepton, and no tracks close to the candidate with more than 2\,GeV for $e$ and 1\,GeV for $\mu$ are allowed. For both isolation types a $\Delta R=0.2$ cone is used. 
Since an Opposite Sign Same Flavour (OSSF) lepton pair can be expected from a neutralino decay, its presence in the event is required in this analysis. Moreover, a cut on $\met>30$\,GeV is applied. Events with $\mossf<20$\,GeV and $81.2\,{\rm GeV}<\mossf<101.2$\,GeV are vetoed in order to reject the $\gamma$ and the $Z$~mass peaks. The resulting $\mossf$ distribution for the benchmark point SU2 and 10\,fb\ is shown in Figure~\ref{fig:1l_bgr}. Due to the clean leptonic signal and the ``soft'' cut on $\met$ this analysis is characterised by a very low systematic uncertainty of around~5\%.

\subsection{Expected mSUGRA Reach at ATLAS and CMS}

\begin{figure}
\begin{center}
\begin{picture}(175,54)
\put(0,0){
\includegraphics[width=8.5cm,height=5.4cm,clip]{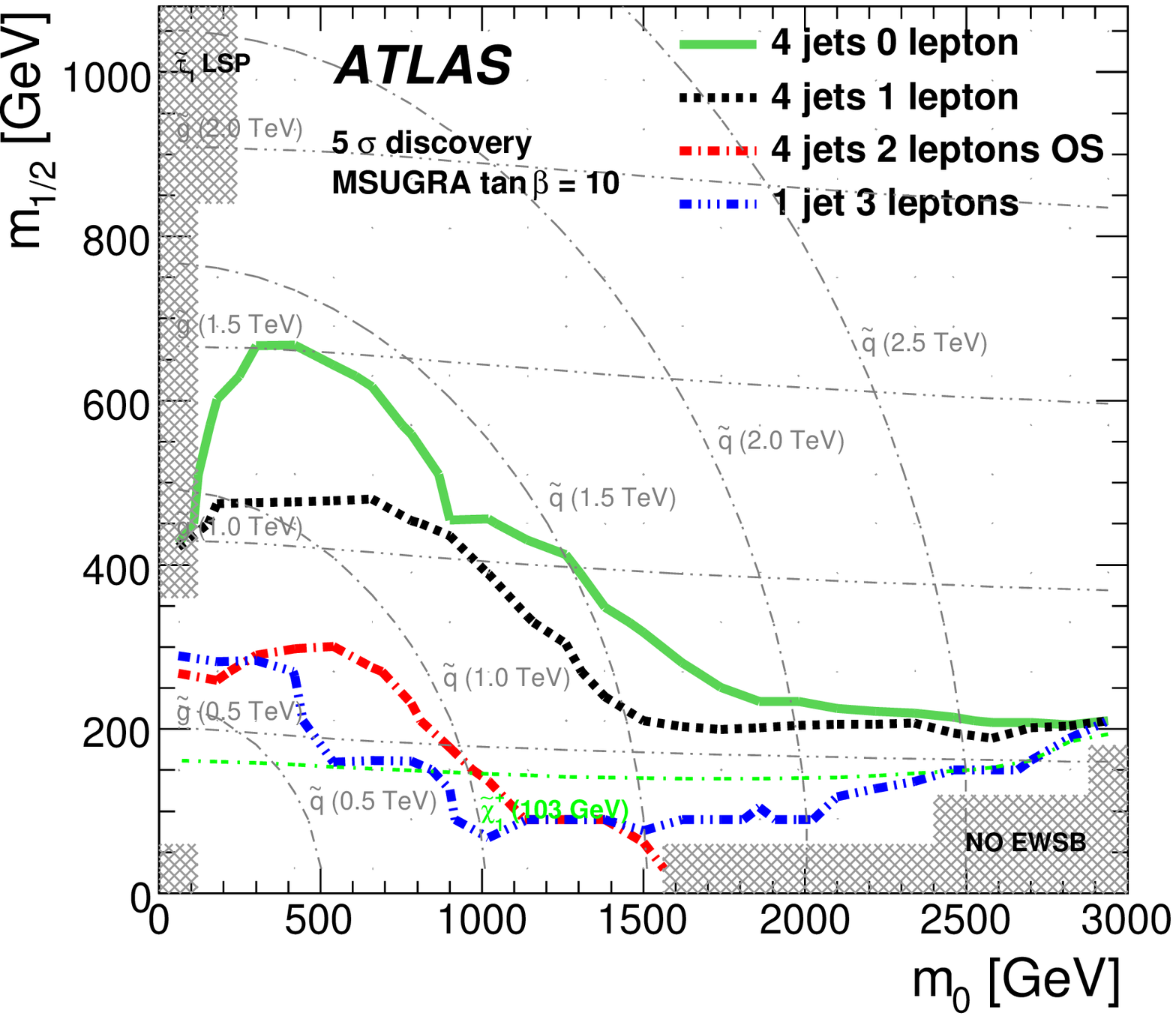}
}
\put(85,1.5){
\includegraphics[width=8.5cm,height=5.3cm,clip]{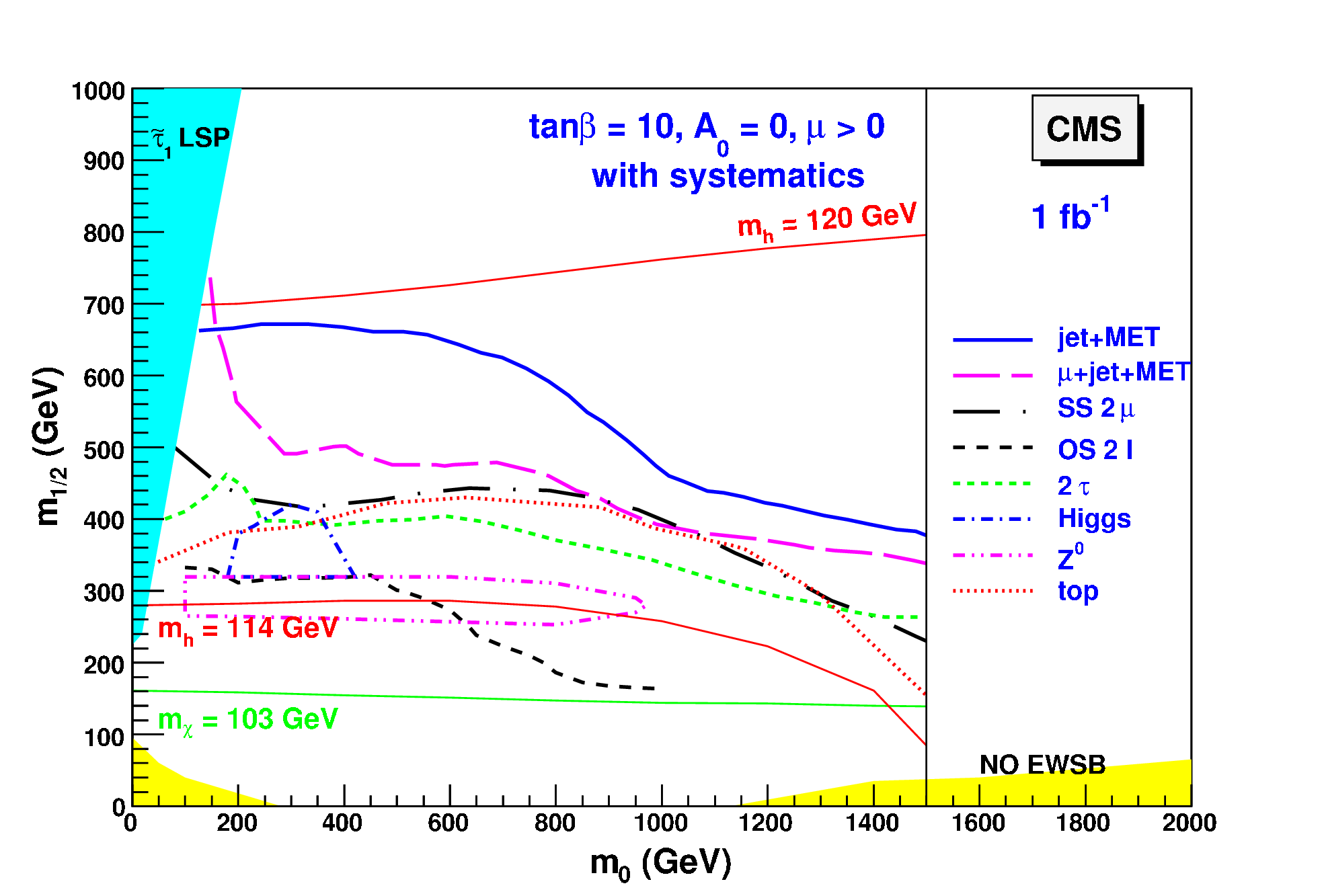}
}
\put(42,48) {($a$)}
\put(113,48) {($b$)}
\end{picture}
\caption{\label{fig:mSUGRAreach}
The expected 5$\sigma$ reach contours including systematic uncertainties for 1\,\fb\ of LHC data for the ($a$)~the ATLAS and ($b$)~the CMS experiments in the mSUGRA $M_0$--\Mhalf\ plane for $\tan\beta=10,~\mu>0,~A_0=0$. See text for details.
}
\end{center}
\end{figure}

The expected discovery reach in the mSUGRA parameter space has been evaluated by both collaborations. The results for $\tan\beta=10,~\mu>0,~A_0=0$ and 1\,\fb\ including systematics are shown as 5$\sigma$ contours in the $M_0$--\Mhalf\ plane in Figure~\ref{fig:mSUGRAreach}. Note that the two collaborations use different statistical significance estimators~\cite{bib:physicsTDR,bib:csc}. Also the systematic uncertainties are treated differently: ATLAS assumes a 20\% uncertainty on the contribution from \ttbar\ as well as vector boson + jet processes, and 50\%  from QCD, while CMS treats its analyses individually. Moreover, in the CMplot the systematic uncertainties are assumed for 10\,\fb\ except for the no-lepton mode (1\,\fb).

It can be concluded, that both experiments have a very similar reach in the mSUGRA parameter space, 
providing for a tight race for BSM physics discoveries!

\section{GMSB Searches with Photons}
Besides generic signatures described in Section~\ref{sec:generic}, both ATLAS and CMS are investigating more exotic final states. In this Section, studies based on high-$p_T$ photon signatures will be discussed. Such final states can for example be produced in GMSB type of models, where the Next-to-Lightest Supersymmetric Particle (NLSP) is a neutralino. It decays to the LSP which is a gravitino, and an energetic photon. Depending on the life time $c\tau$ of the NLSP, photons can be either \textit{pointing} or \textit{non-pointing} to the primary vertex, which determines the search strategy.

\subsection{Search for GMSB Models with Photon Signatures at CMS}
The CMS collaboration has set up a search analysis for GMSB using high-$p_T$ photons~\cite{bib:zalewski}. The six GMSB parameters defining the signal benchmark are summarised in~\cite{bib:parametersCMSGMSB,bib:zalewski}. Two scenarios 
will be discussed here: pointing and non-pointing photons, with an NLSP decay length of $c\tau=0$ and 100\,cm, respectively.

\begin{figure}
\begin{center}
\begin{picture}(175,54)
\put(14,0){
\includegraphics[width=7cm,height=5.4cm,clip]{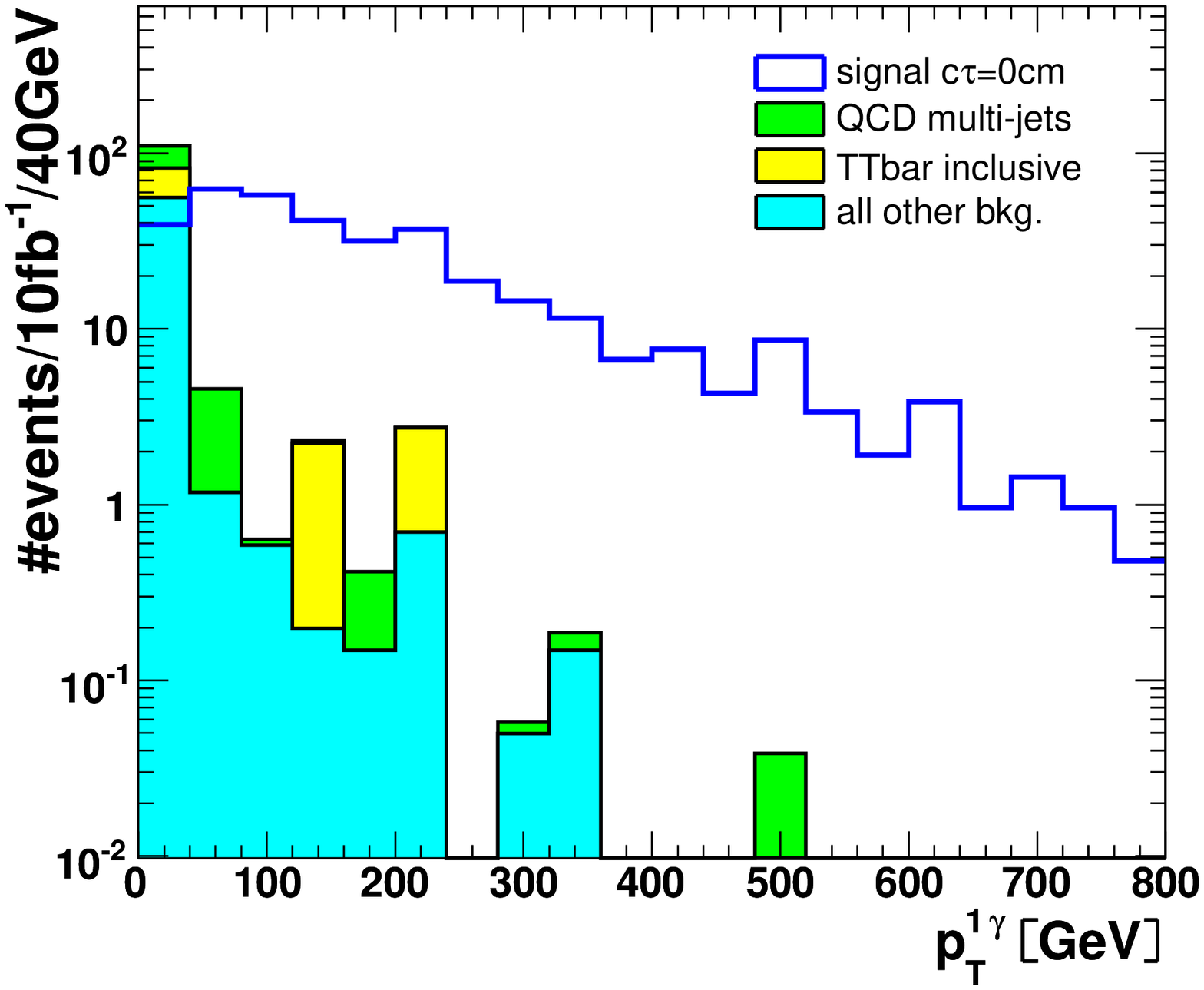}
}
\put(83,0){
\includegraphics[width=7cm,height=5.4cm,clip]{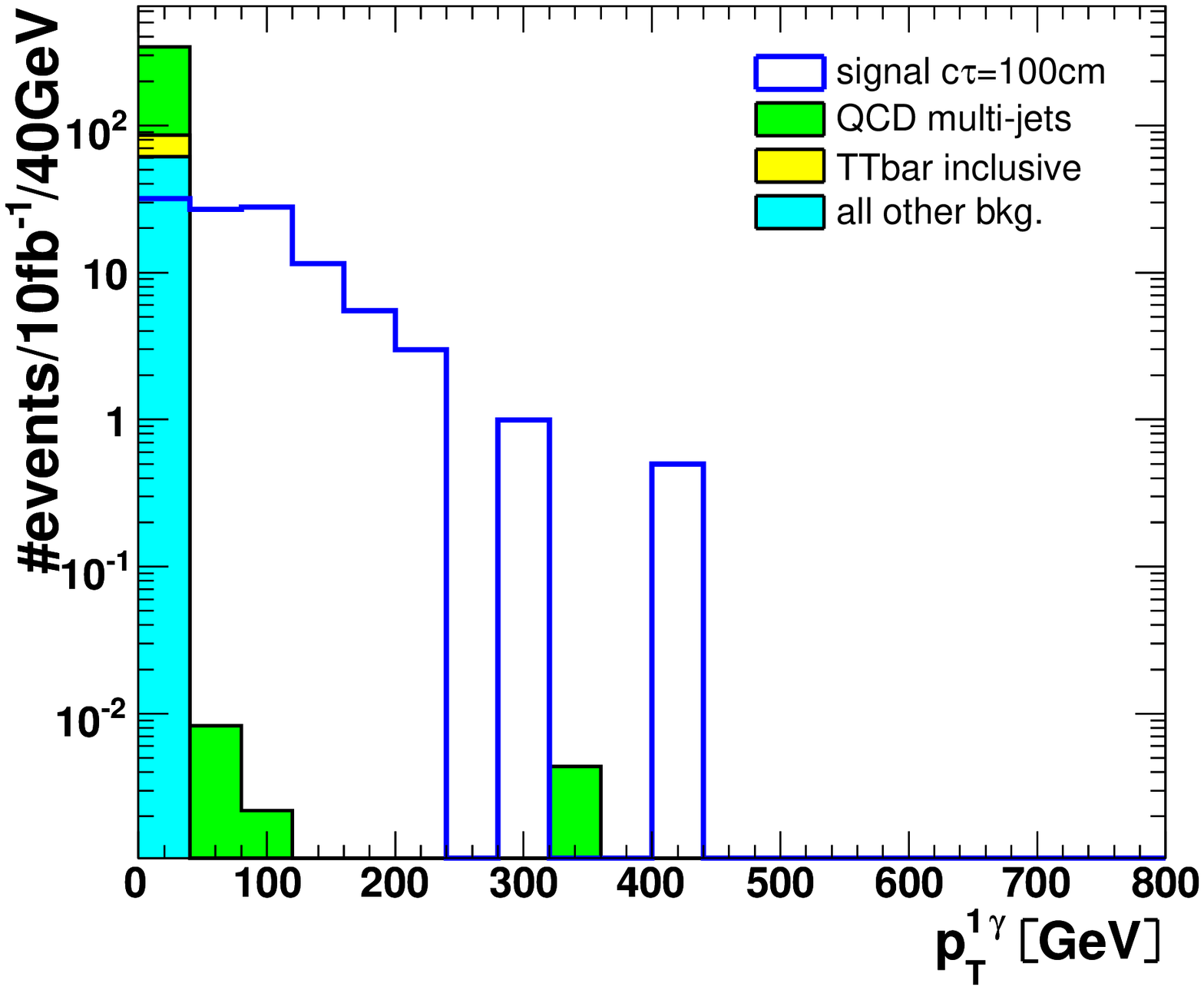}
}
\put(30,48) {($a$)}
\put(99,48) {($b$)}
\end{picture}
\caption{\label{fig:zalewski}
Distribution of the leading photon transverse momentum $p_T^{\gamma_1}$ after all cuts except those on $p_T^{\gamma_1}$ itself for ($a$)~pointing and ($b$)~non-pointing (NLSP decay length $c\tau=100$\,cm) scenarios. No trigger requirement is applied.}
\end{center}
\end{figure}

This analysis triggers on high transverse momentum photons with $p_T^\gamma>80$\,GeV. In the offline selection, at least one photon passing the same transverse momentum threshold is required. In order to suppress backgrounds from QCD, this photon is to be track-isolated: the algebraic sum of $p_T$'s of tracks in a $\Delta R<0.3$ cone around the photon candidate should not exceed 3\,GeV. The high-level trigger is less efficient for the non-pointing case (17\%) than for the pointing case (23\%).

Beyond that, a selection similar to those presented in Section~\ref{sec:generic} is made: four jets are required in $|\eta|<3$ with $p_T^j>50$\,GeV, of which the leading one needs to be central: $|\eta^{j_1}|<1.7$. For missing transverse energy, which should not be aligned with any of the four leading jets: $\phi(\met,p_T^{j_k})>20^o,\,k=1,...4$, at least $\met>160$\,GeV is required.

Additional sophisticated cuts are applied on the photons. Shower shape variables based on the transverse energy covariance tensor
\begin{equation}
  C_{\eta,\phi} \equiv 
   \left( \begin{array}{lr}
   S_{\eta\eta} & S_{\eta\phi} \\
   S_{\phi\eta} & S_{\phi\phi} \\
   \end{array} \right)
  ,\quad{\rm where}~~
  S_{\alpha\beta} \equiv \frac1{E^\gamma}\cdot{\sum}_i E_i^{\rm cell} \cdot 
   (\alpha_i-\langle\alpha\rangle) (\beta_i-\langle\beta\rangle)
  ~~{\rm and}~~i\in\{\rm all~cells~of~\gamma~cluster\}
\end{equation}
are formed, and cut sets optimised for the pointing and non-pointing case are defined. This part of the analysis is detailed in~\cite{bib:zalewski}.

After all cuts, the resulting photon momenta distributions shown in Figure~\ref{fig:zalewski}, which demonstrate a remarkable signal-to-background ratio.

\subsection{Expected Reach for GMSB with Photon Signatures at ATLAS and CMS}
\begin{figure}
\begin{center}
\begin{picture}(175,54)
\put(10,0){
\includegraphics[width=7cm,height=5.4cm,clip]{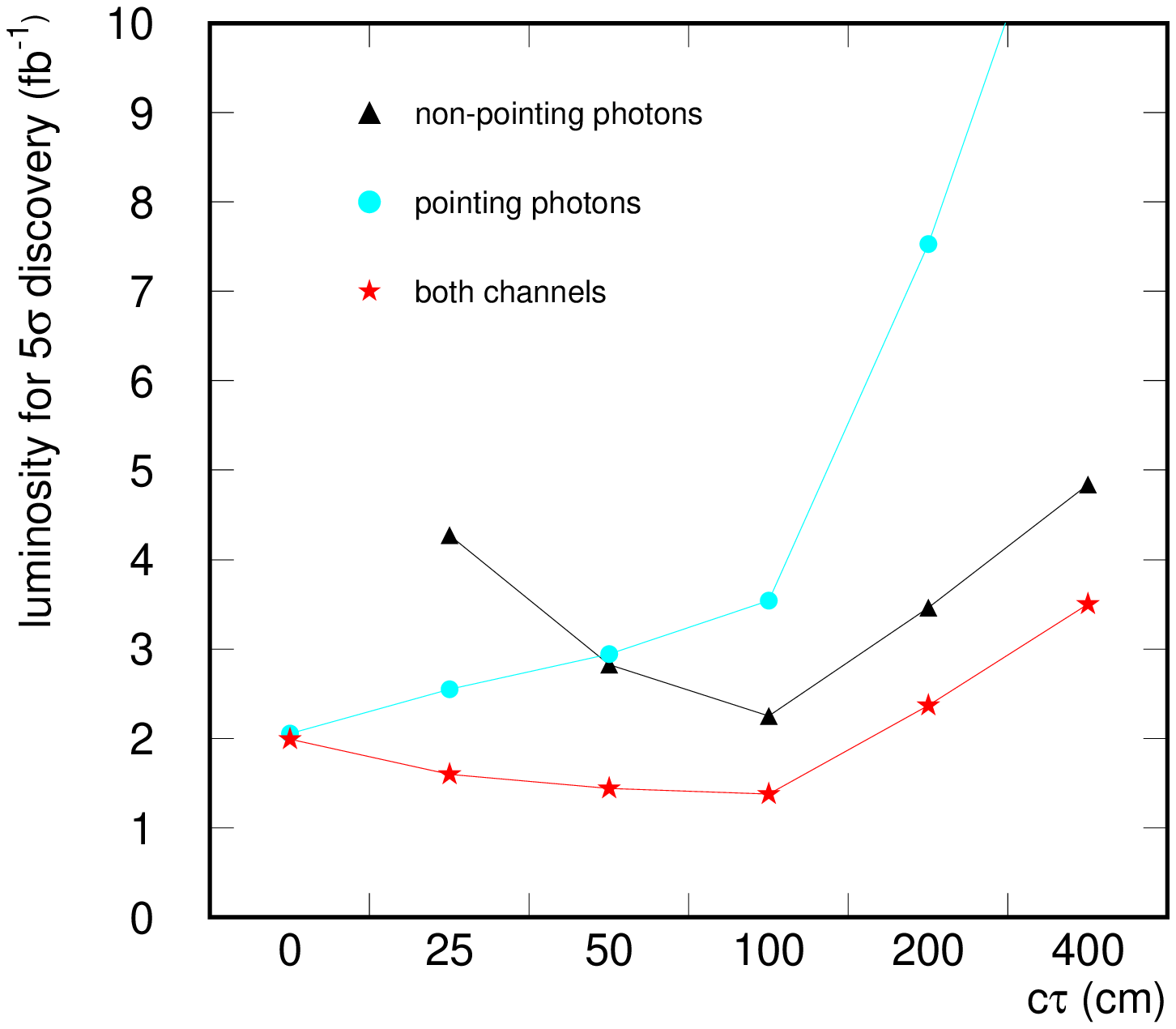}
}
\put(85,0){
\includegraphics[width=7cm,height=5.4cm,clip]{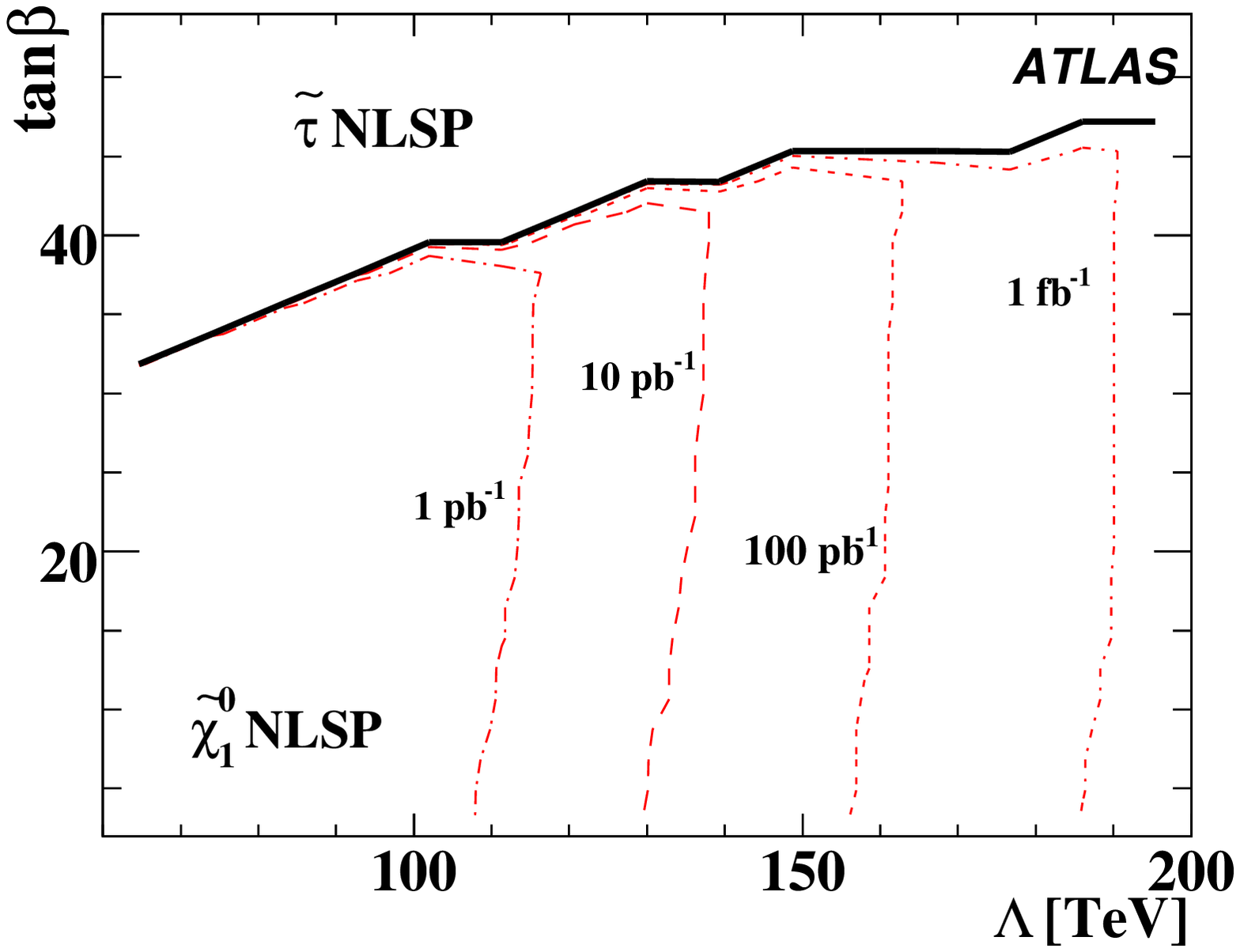}
}
\put(27,48) {($a$)}
\put(95,48) {($b$)}
\put(36,32) {{\sf\large CMS}}
\end{picture}
\caption{\label{fig:reachGMSB}
($a$)~Integrated luminosity needed by CMS for a 5$\sigma$ discovery of GMSB with various NLSP decay lengths $c\tau$. ($b$)~Expected reach of ATLAS with pointing photon signatures as 5$\sigma$ contours in the $\Lambda$-$\tan\beta$ plane. For details about the GMSB models used by ATLAS and CMS for these studies see text.
}
\end{center}
\end{figure}

The integrated luminosity needed for a 5$\sigma$ discovery by the CMS GMSB search analysis discussed in the previous Subsection is presented in Figure~\ref{fig:reachGMSB}~($a$) for various NLSP decay lengths: $c\tau=0,\,25,\,50,\,100,\,200,\,400$\,cm. As expected, the search based on pointing signatures fares better in scenarios where $c\tau\rightarrow0$ than for high $c\tau$ values. Non-pointing photon search performance reaches its maximum for an NLSP decay length of $c\tau\simeq100$\,cm and deteriorates beyond, as fewer NLSP's decay \textit{inside} of the tracker volume. The ATLAS experiment also investigates pointing and non-pointing photonic signatures~\cite{bib:csc}. The reach of ATLAS with pointing photon signatures in the $\Lambda$-$\tan\beta$ plane is shown in Figure~\ref{fig:reachGMSB}~($b$) for a set of GMSB parameters summarised in~\cite{bib:parametersATLASGMSB}. The statistical significances used by the two collaborations are documented in~\cite{bib:zalewski,bib:csc}, respectively.

\section{Leptonic Long-Lived Massive Particles and $R$-Hadron Searches}
Leptonic massive particles which are stable on the time scale needed for detection in collider experiments are predicted in many BSM models. One example is GMSB with $N_{\rm gen5}>1$ and $\tan\beta\gg1$, so that a slepton is the NLSP and couples weakly to the gravitino. A variety of such scenarios are investigated at ATLAS~\cite{bib:csc} and CMS~\cite{bib:physicsTDR}.

\subsection{Search for Leptonic Long-Lived Massive Particles at ATLAS}
\begin{figure}
\begin{center}
\begin{picture}(175,42)
\put(0,0){
\includegraphics[height=4cm,width=5.4cm,clip]{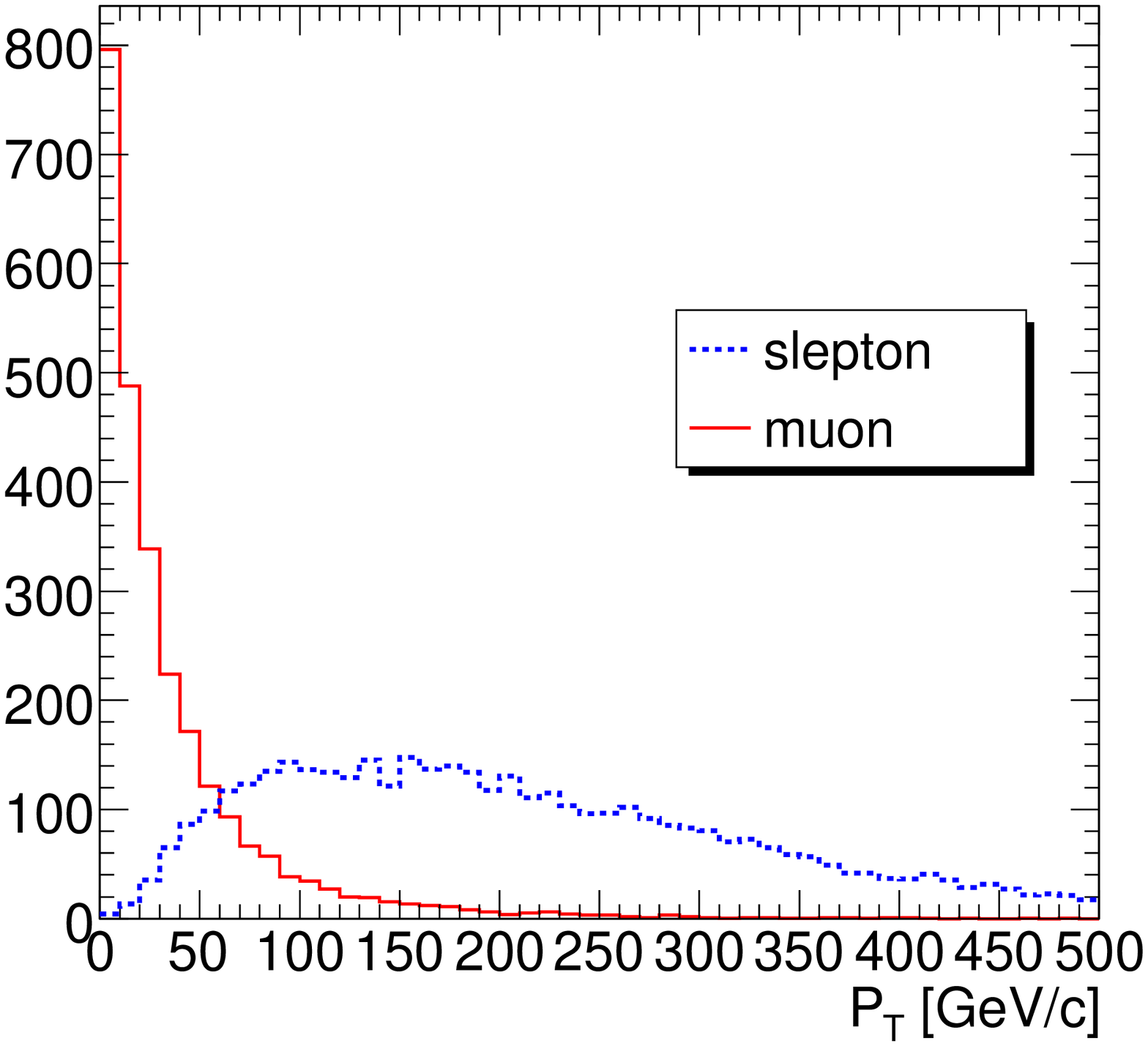}
}
\put(56,0){
\includegraphics[height=4cm,width=5.4cm,clip]{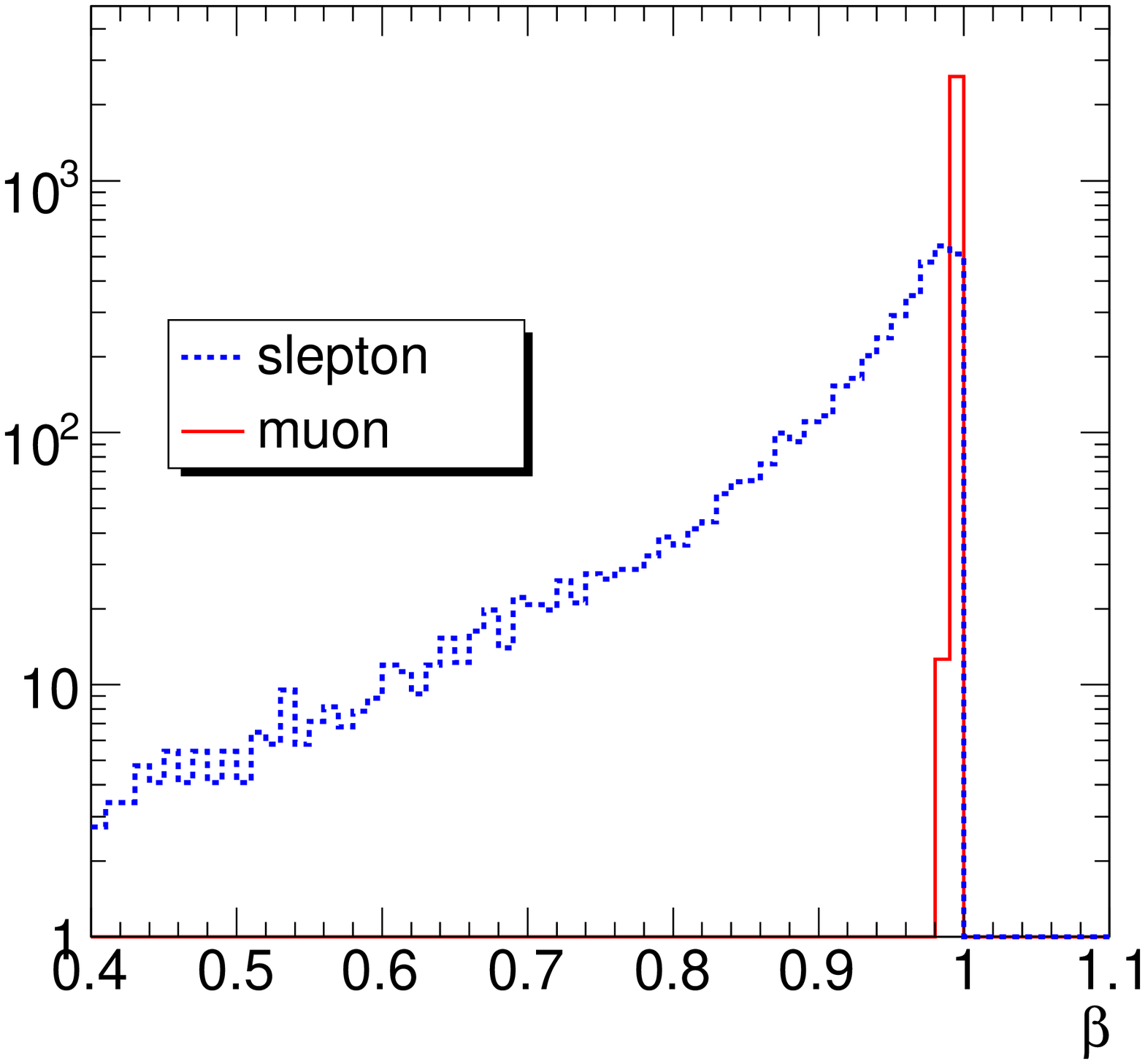}
}
\put(116,1){
\includegraphics[height=3.9cm,width=5.2cm,clip]{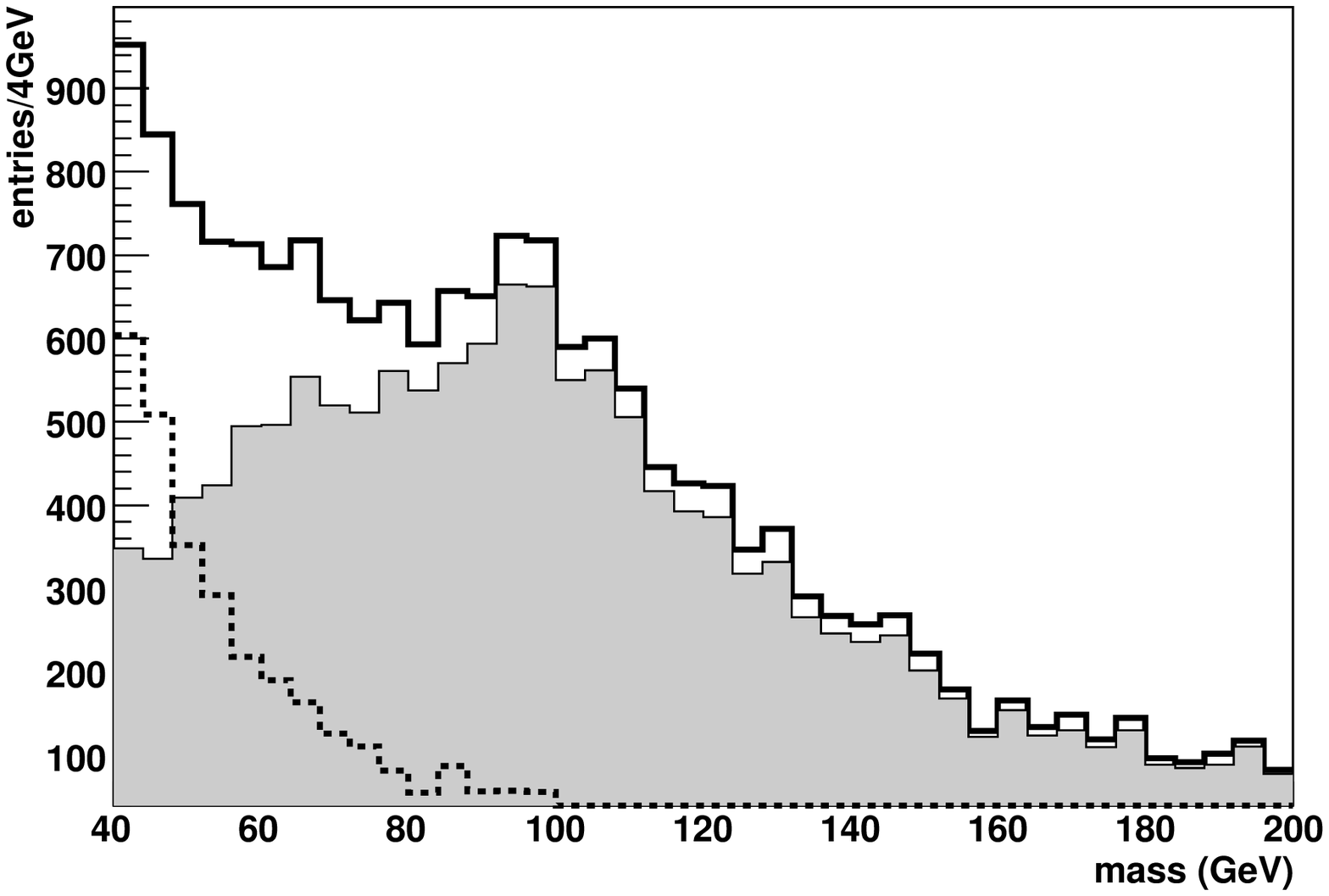}
}
\put(10,35) {($a$)}
\put(66,35) {($b$)}
\put(126,35) {($c$)}
\end{picture}
\caption{\label{fig:slepton}
For the ATLAS search of massive long-lived leptonic particles ($a$)~reconstructed $p_T$ and ($b$)~velocity $\beta$ are shown on arbitrary scale for muons and sleptons. ($c$) Mass distribution of level-2 trigger candidates for 0.5\,\fb. The shaded histogram represents massive long-lived sleptons, and the dashed open one muons. The solid open histogram is their total.
}
\end{center}
\end{figure}

Leptonic long-lived massive particles offer a distinct signature. As they are stable, they produce a penetrating track similar to muons. However, they are significantly less boosted than muons due to their high mass. Typically, their velocity peaks at $\beta\simeq1$, but has a wide tail towards $\beta\ll1$. The $p_T$ and velocity spectra of leptonic NLSPs and muons is shown in Figure~\ref{fig:slepton}. This and all other figures in this Subsection are for GMSB parameters summarised in~\cite{bib:parametersATLASWIMP}.

A major challenge in searches for leptonic long-lived massive particles is the size of the ATLAS detector. With a radius of roughly 10\,m, signals from three bunch crossings coexist in parts of the detector at the same time. The read out chain is optimised for ultra relativistic particles with $\beta\rightarrow1$, and a sharp drop in trigger efficiency is observed at a velocity of $\beta\simeq0.8$. Therefore it is highly important to access the information from neighbouring bunch crossings upon a trigger signal. Such a possibility is implemented in the ATLAS DAQ for major components of the Muon Spectrometer for at least $\pm$1 bunch crossing~\cite{bib:csc}.

In this study, a high-$p_T$ muon trigger at level-1 was found to be 95\% efficient. At level-2, timing information from the resistive plate chambers in the barrel part of the muon spectrometer~\cite{bib:tdr} can be used. Requiring $p_T>40$\,GeV, $\beta<0.97$ and $M>40$\,GeV, a trigger efficiency of about 50\% can be achieved at level-2. The resulting distribution of the reconstructed slepton mass at level-2 for 0.5\,\fb{} is shown in Figure~\ref{fig:slepton} ($c$) for an input mass of 100\,GeV. 

\begin{figure}
\begin{center}
\begin{picture}(175,45)
\put(18,0){
\includegraphics[width=6.3cm,height=4.7cm,clip]{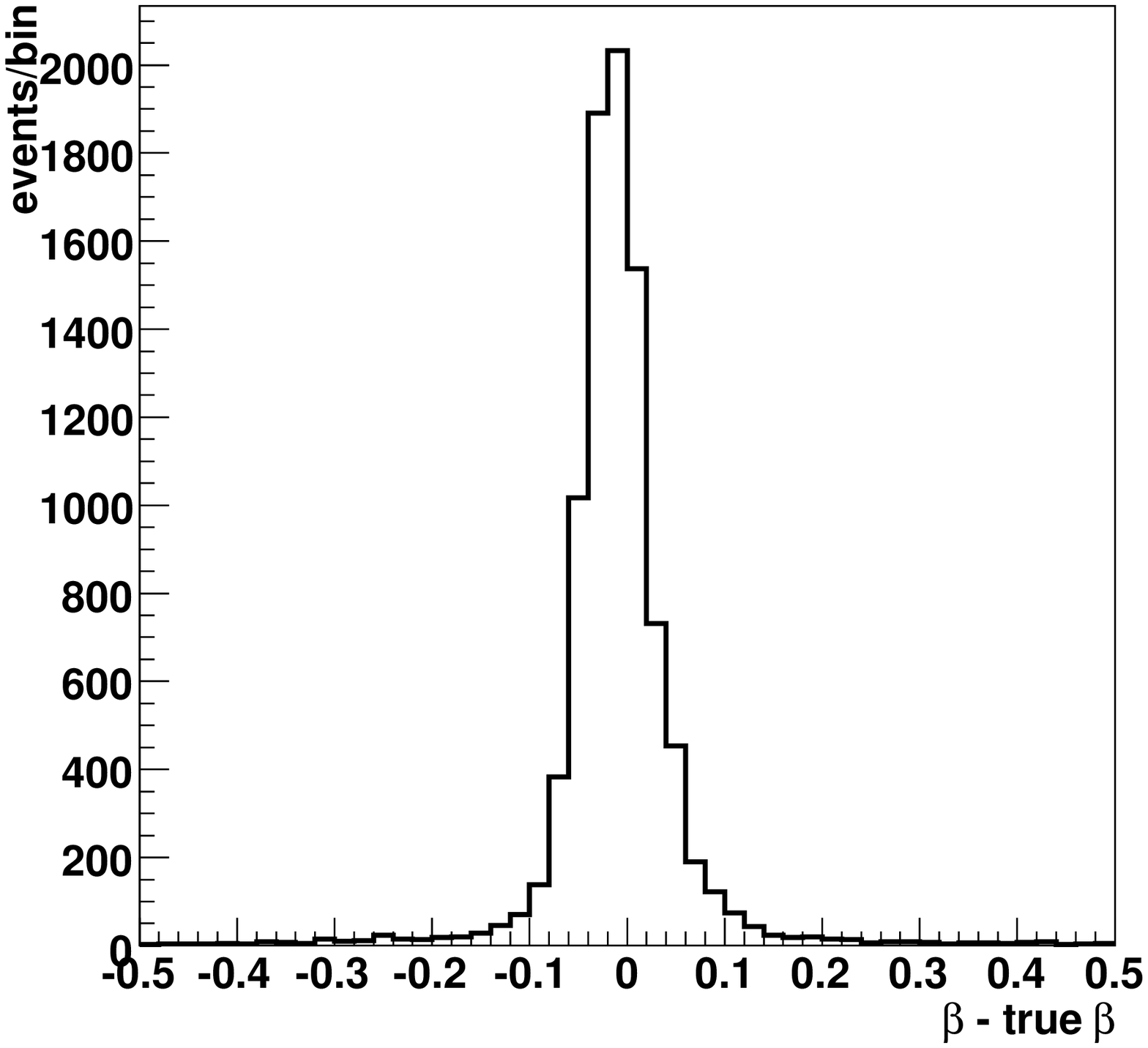}
}
\put(86,0){
\includegraphics[width=6.3cm,height=4.7cm,clip]{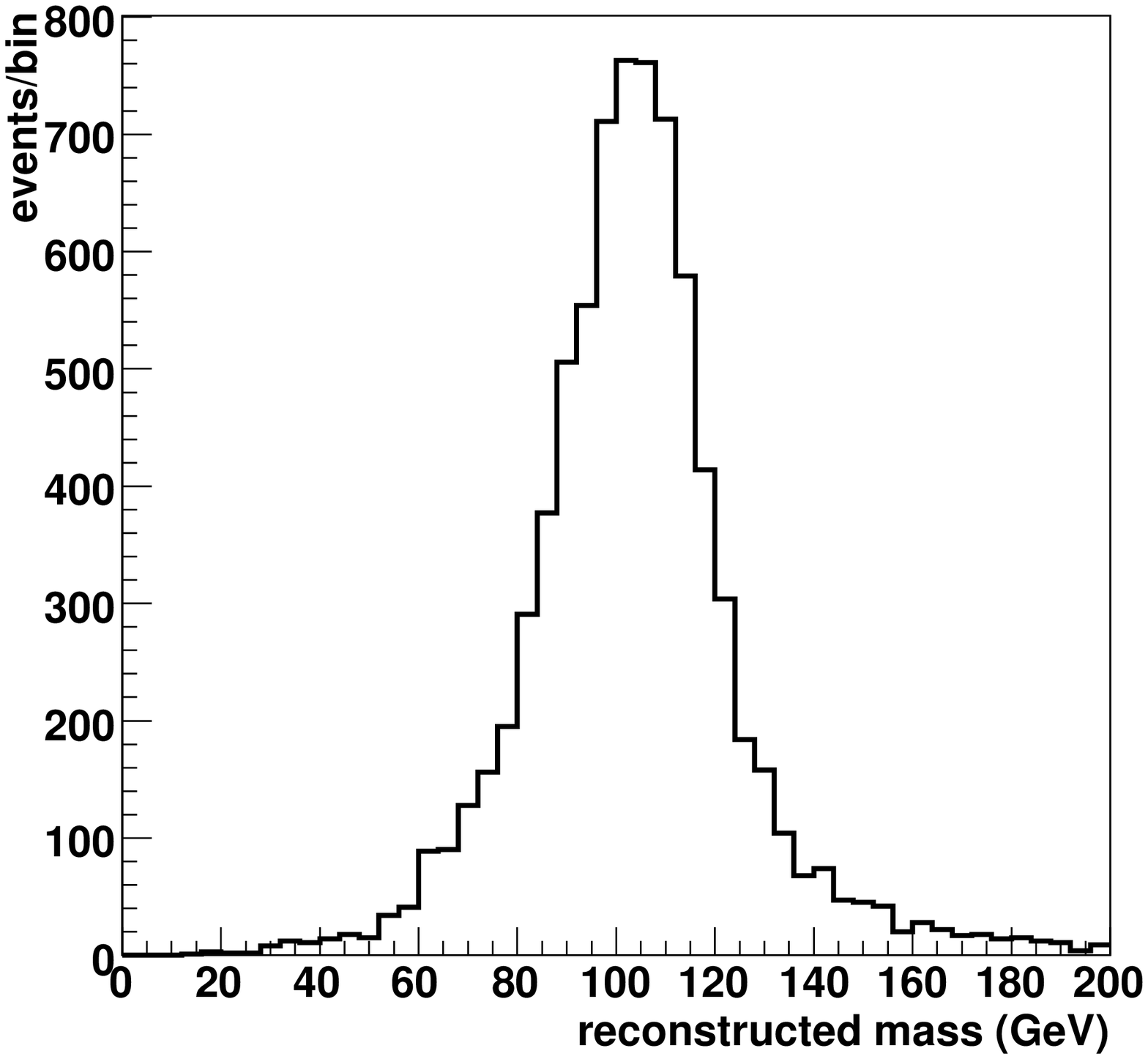}
}
\put(30,42) {($a$)}
\put(97,42) {($b$)}
\end{picture}
\caption{\label{fig:sleptonReco}
($a$)~The velocity resolution and ($b$)~the reconstructed slepton mass are shown for the ATLAS search of massive long-lived leptonic particles. The simulated slepton mass is 100\,GeV. The scale is arbitrary.
}
\end{center}
\end{figure}

The offline selection of long-lived massive sleptons pursues the same strategy as the level~2 trigger. For their reconstruction the $\chi^2$ of the track fit is minimised not only with respect to the associated hit positions, but also the time of arrival. The resulting momentum resolution is shown in Figure~\ref{fig:sleptonReco}~($a$), and the resulting mass distribution for an input mass of~100\,GeV in~($b$).

\subsection{$R$-Hadron Searches at ATLAS and CMS}
The experimental challenge is even bigger for stable long-lived hadronic particles, also known as $R$-hadrons, which are predicted in many models, see references in~\cite{bib:csc}. They hadronise as they are traversing the detector, and typically undergo $\mathcal O(10)$ nuclear interactions. In these interactions they can change their charge in multiples of 1/3, and even become neutral. Therefore searches at ATLAS and CMS involve a variety of signatures based on track stubs in the Inner Detector and the Muon Spectrometer, charge-flipping tracks and many more, which are described in~\cite{bib:csc, bib:physicsTDR}.

\section{Conclusion and Outlook}
It can be concluded that ATLAS and CMS have investigated a wide variety of signatures expected from supersymmetric BSM physics. Both collaborations prepared inclusive and specialised search analyses, of which only a small fraction can be presented in these proceedings. The experimental challenges of early data analysis have been critically confronted, and various background estimation techniques from data developed. The way is paved towards first physics mode collisions at a centre-of-mass energy of $\sqrt s=10$\,TeV, which will hopefully happen later this year. Both collaborations are eagerly awaiting to investigate the physics at TeV scale. 

\begin{acknowledgments}
I would like to express my appreciation to all of my colleagues working on SUSY at ATLAS and CMS for their progress in the last two years. I would like to thank Pawel Br\"uckman de Rentstrom, Alan Barr, M\"uge Karag\"oz \"Unel, Tony Weidberg, David Costanzo, Giacomo Polesello, Sarah Eno, and Albert de Roeck for their vital help in preparing this talk.
\end{acknowledgments}


\end{document}